\documentclass{article}

\usepackage[numbers,sort&compress]{natbib}
\usepackage{style}
\usepackage{fancyhdr}
\fancypagestyle{empty}{\fancyhf{}\fancyfoot[L]{\begin{minipage}[b]{0.45\textwidth}\rule{2in}{0.4pt}\\[2pt]\footnotesize $^\dagger$Equal contribution. Under review.\end{minipage}}\fancyfoot[C]{\thepage}}

\usepackage[utf8]{inputenc}
\usepackage[T1]{fontenc}
\usepackage{hyperref}
\usepackage{url}
\usepackage{booktabs}
\usepackage{amsfonts}
\usepackage{nicefrac}
\usepackage{microtype}
\usepackage{xcolor}
\usepackage{colortbl}

\usepackage{amsmath}
\usepackage{algorithm}
\usepackage{algpseudocode}

\usepackage{wrapfig} 
\usepackage{orcidlink}

\makeatletter
\renewcommand*{\@fnsymbol}[1]{\ifcase#1\relax\dagger\or\ddagger\or\S\or\P\or
    \|\or **\or \dagger\dagger \or \ddagger\ddagger \else\@ctrerr\fi}
\makeatother

\newsavebox{\panelbox}
\newcommand{\panel}[2]{\leavevmode\sbox{\panelbox}{#2}\llap{\raisebox{\dimexpr\ht\panelbox-\height\relax}{\small\textbf{#1}}\hspace{0.5em}}\usebox{\panelbox}}

\title{Solving Stochastic Inverse Problems \\ in Embedding Space}

\author{%
  Elias Najarro$^\dagger$ \\
  \orcidlink{0000-0002-7875-3251} \href{https://orcid.org/0000-0002-7875-3251}{0000-0002-7875-3251} \\
  IT University of Copenhagen\\
  2300 Copenhagen, Denmark \\
  \And
  Nicolas Bessone$^\dagger$ \\
  \orcidlink{0009-0008-9008-0003} \href{https://orcid.org/0009-0008-9008-0003}{0009-0008-9008-0003} \\
  IT University of Copenhagen\\
  2300 Copenhagen, Denmark \\
  \And
  Sebastian Risi \\
  \orcidlink{0000-0003-3607-8400} \href{https://orcid.org/0000-0003-3607-8400}{0000-0003-3607-8400} \\
  IT University of Copenhagen \\
  2300 Copenhagen, Denmark \\
}

\begin{document}

\maketitle

\begin{abstract}

Solving an inverse problem consists in finding the unknown parameters of a system from observations. This task is particularly challenging when observations have a strong stochastic component, producing diverse yet equivalent patterns. Self-organizing systems are a notable case: simple local rules can generate complex stochastic patterns. Many natural systems rely on such processes, making self-organization central to understanding natural phenomena. Inverse methods based on direct comparison of observed values struggle in this setting, as such comparisons cannot capture feature similarities between variable outcomes; as a result, the loss fails to meaningfully inform the optimizer. In this work, we show that the difficulty posed by stochasticity in observable patterns can be overcome by reformulating the inverse problem in an embedding space. By mapping stochastic patterns onto an invariant embedding space, where equivalent patterns have similar representations, we can effectively recover unknown causal parameters without the need for handcrafted objective functions or heuristics. We show that embedding models trained on large image datasets produce robust representations that capture these invariances. We evaluate the method on three self-organizing systems: a physical, a biological, and a social one; namely, a reaction-diffusion system, a model of embryonic development, and an agent-based model of social segregation. We show that the method reliably recovers parameters despite stochasticity in the pattern outcomes. We further apply the method to real biological patterns, highlighting its potential as a tool for both theorists and experimentalists to investigate the dynamics underlying complex stochastic pattern formation.

\end{abstract}

\section{Introduction}

\begin{figure}[H]
    \centering
    \panel{a}{\raisebox{-3pt}{\includegraphics[height=0.29\textwidth, trim=9.5 4.5 9.8 5.5, clip]{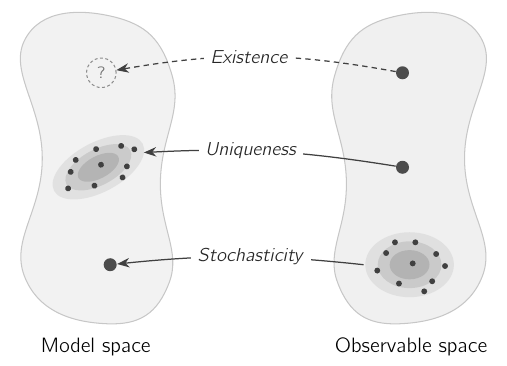}}}\hspace{0.07\textwidth}%
    \panel{b}{\includegraphics[height=0.29\textwidth, trim=0 1 0 7.8, clip]{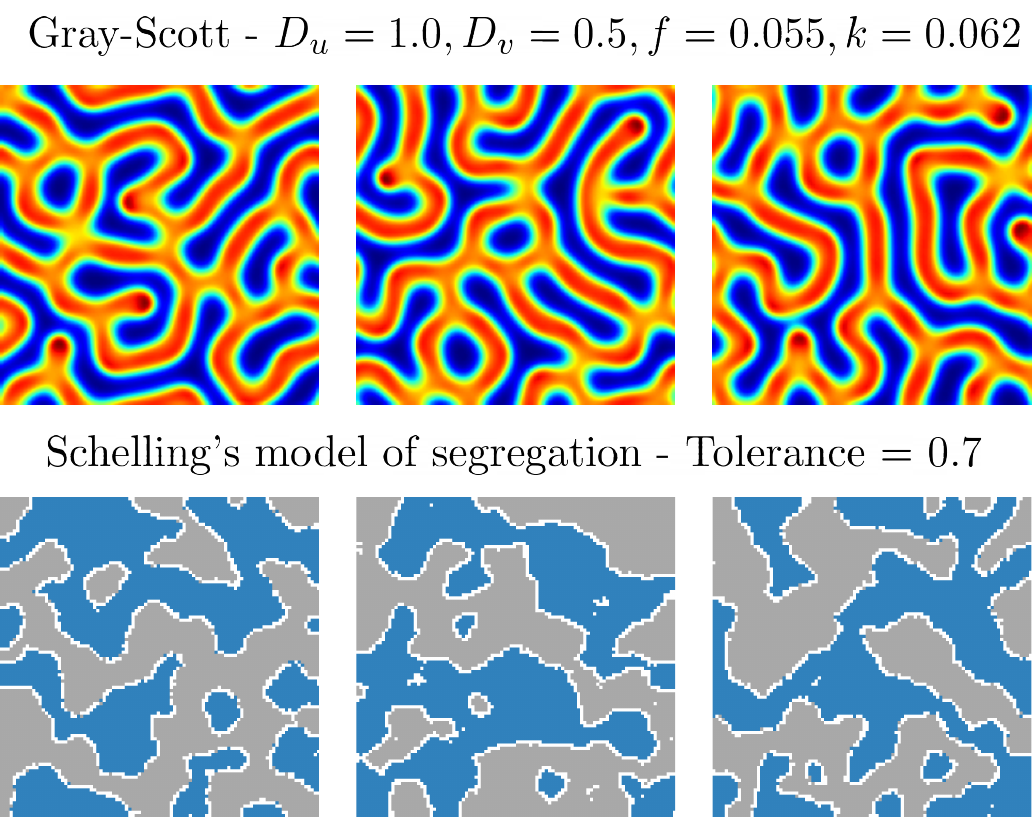}}
    \caption{\textbf{Inverse problems and stochasticity in self-organizing systems.} (\textbf{a}): An \textit{\textbf{inverse problem}} consists in finding the mapping from observations to their model space (also referred to as \textit{parameter space}, \textit{solution space} or \textit{domain space} \cite{tarantola_inverse_2005, bunge_inverse_2019}). It presents three challenges: (i) existence: solutions may not exist for all observations; (ii) uniqueness: multiple causes may produce identical observations; and (iii) stochasticity: observations may differ for the same model parameters. (\textbf{b}): Three simulation instances of two self-organizing models run with the same parameters.~\textit{(Top)}: Reaction-diffusion model; the stochasticity results from varying initial conditions $s_0\sim\mathcal{S}$.~\textit{(Bottom)}: Schelling's segregation model; the stochasticity results from asynchronous updates in the model $F$. Both are examples of stochasticity in the observable space, the challenge our method addresses.}
    \label{fig:splash}
\end{figure}

Inverse modeling addresses the problem of inferring unknown model parameters from observations: ``\textit{Given these macroscopic observations, what model parameters produce them?}'' \cite{tarantola_inverse_2005, bunge_inverse_2019}. An inverse model takes an observation as input and returns the model, or initial configuration, that generates the observation. See Fig. \ref{fig:splash}a for a visual summary of inverse modeling and some of its difficulties.

A major challenge in inverse problems is the presence of stochasticity in the observations. Natural phenomena, notably those associated with complex systems, exhibit sensitivity to initial conditions. Differences in initial conditions or stochastic dynamics can produce different observable patterns even when the underlying parameters remain fixed \citep{durrett1999stochastic, murray2007mathematical, ball2009shapes}. Often, these patterns share high-level features but do not match when compared pixel by pixel. They are stochastic instances of the same class of patterns (Fig.~\ref{fig:splash}b). For instance, while every human fingerprint is unique, they all share characteristic features that make them recognisable as fingerprints. Similarly, no two leopards have identical spot motifs, yet all are immediately identifiable as leopard patterns. In mathematical terms, patterns can belong to the same class in the presence of invariances--be it translational, rotational, or of another kind.

Existing inverse modeling approaches address stochasticity in the model space \cite{Vauhkonen2016Jul, Ye2019Mar, kawar2021snips}. Methods for stochasticity in the observable space, such as approximate Bayesian computation (ABC), rely on summary statistics that must be hand-designed or learned for each system \cite{jiang2017learning}.

In this work, we introduce a method that solves inverse problems in embedding space, addressing the challenge of stochasticity in observable patterns. The method operates by mapping target patterns onto an invariant representation, where patterns generated by the same parameters from different initial states have similar representations, and combines these representations with evolutionary strategies to search the parameter space. We demonstrate that such a representation can be built both through a multi-scale wavelet decomposition and with visual embedding models trained on large image datasets, whose emergent representations naturally capture invariances useful for recovering the parameters across systems. This enables efficient recovery of causal parameters in stochastic self-organizing systems without the need for heuristics or hand-crafted loss functions. The method works across systems without having to learn specific statistics or metric spaces for each system.

We evaluate the method on three distinct--physical, biological, and social--self-organizing systems: Gray-Scott's reaction-diffusion model, a model of early embryonic development in mammals, and an agent-based model of social segregation. These systems provide examples of stochasticity in the observable space: random initial conditions or stochastic dynamics produce patterns that differ at the microscopic level while sharing macroscopic features for the same model parameters. We compare parameter recovery across three embedding models and five non-learned similarity metrics. Results show that the method recovers parameters across these three systems despite stochasticity in the observable patterns.

\section{Background}

\vspace{-1mm}
\paragraph{Inverse problems}
\label{inverse problems}
Finding the solution to an \emph{inverse problem} consists in determining the unknown parameters, or initial conditions, of a system from observations of its outcomes~\cite{tarantola_inverse_2005}. By contrast, \textit{forward problems} unambiguously map inputs to outputs using the system model. Formally, solving an inverse problem requires finding $(\theta, s_0) \in \Theta \times S$ such that $F(\theta, s_0) = y_{\text{obs}}$, where $F: \Theta \times S \to Y$ denotes the forward model mapping parameters $\theta \in \Theta$ and initial states $s_0 \in S$ to observed data $y_{\text{obs}} \in Y$. Inverse problems are typically ill-posed, characterised by solution sensitivity and non-uniqueness \cite{Groetsch1993}. These properties link inverse problems to complex systems: small changes in their parameters or initial state can result in divergent outcomes. As such, recovering the causal parameters--i.e., solving the inverse problem--shares similar challenges with ill-posed problems.

We use model space to encompass functional forms, parameters, and initial conditions. Stochasticity can be present in inverse problems in two distinct ways. \textbf{Stochasticity in the model space}: multiple parameters or initial conditions $(\theta, s_0) \in \Theta \times S$ map to identical or nearly identical observations $y_{\text{obs}} \in Y$, resulting in degenerate solutions and non-uniqueness, i.e., non-injectivity of $F$. In this case, the unknown quantities in the model space are distributions. An example is seismic tomography, which uses seismic waves recorded at the surface to reconstruct the Earth's interior. Solutions are non-unique: similar surface measurements can originate from different inner structures. \textbf{Stochasticity in the observable space}: in this case, observations are inherently stochastic. Either resulting from randomness in initial states, such as in reaction-diffusion systems ($y_{\text{obs}} = F(\theta, s_0)$ with $s_0 \sim \mathcal{S}$), or from intrinsic stochasticity in the forward model $F$, as in the asynchronous updates of an agent-based model. See Fig. \ref{fig:splash}b. Stochasticity in the model space, where similar observations can have different causal origins, has been studied in inverse modeling \cite{Vauhkonen2016Jul, Ye2019Mar, li2020nett, kawar2021snips, xu2021solving, genzel2022solving}. Stochasticity in the observable space has been addressed through simulation-based inference, including approximate Bayesian computation (ABC), which compares simulated and observed data through summary statistics \cite{beaumont2002approximate, cranmer2020frontier}. However, these summaries must be specified for the inference problem. Learning them from simulations automates their construction, but still produces a representation fitted to the system under investigation \cite{jiang2017learning}. Our method avoids having to design or learn these summaries for each system, using embedding representations to guide parameter recovery. The method returns a point estimate of the parameters, without estimating their posterior distribution.

\vspace{-1mm}
\paragraph{Embedding Representations} An embedding maps a pattern to a vector whose distance to other vectors reflects the similarity between patterns. It can be built from a fixed multi-scale wavelet decomposition, as in Fréchet Wavelet \cite{Veeramacheneni2023Dec}, or learned by a visual embedding model. Visual embeddings are learned vector representations that capture relevant features of images beyond the raw pixel values. Instead of encoding only low-level information, such as colour and intensity, embeddings represent higher-level visual attributes, such as shape, texture, structure, and semantic content \cite{frome2013devise, norouzi2013zero, Jia2021Feb,Niklasson2021Feb,Tian2021Sep}. Visual embedding models transform image data into a lower-dimensional space where semantically or visually similar images are mapped to nearby points, while dissimilar images are placed farther apart \citep{Moskvyak, Radford2021Feb, Kumar2024Dec}. See diagram in Fig. \ref{fig:method}b.%

\textit{Geometric invariant representations.} Embedding representations have been shown to be invariant under different pattern transformations \cite{goodfellow2009measuring, zhu2016deep, tschannen2018recent,Zhang2018Jan, ilse2020diva}. For instance, \citet{Moskvyak} showed that convolutional embeddings can produce invariant representations of animal patterning under homographic transformations. We evaluate three embedding models, CLIP \cite{Radford2021Feb}, Nomic \cite{Nussbaum2024Jun} and DINO \cite{caron2021emerging}. Their embeddings are 512-dimensional for CLIP and 768-dimensional for Nomic and DINO. Trained on large-scale image datasets, these models capture high-level visual features, making them suitable for general-purpose applications without requiring additional fine-tuning. Fig. \ref{fig:method}c shows that embeddings are able to capture relevant visual features.

\paragraph{Self-organizing systems}

Self-organization, understood as the spontaneous emergence of global order or coordination in a system from local interactions of its parts, is observed across physical, biological, and social systems \cite{haken1987synergetics, kauffman1992origins, Paczuski1999Jun}. Examples range from morphogenesis, where cells divide and specialise to form complex body plans in living organisms \cite{vonneumann1966theory, Weiss1960Sep}, to animal skin patterns \cite{murray2007mathematical} or flocking birds \cite{reynolds1987flocks}. In social systems, ABMs such as Schelling's segregation model show how simple local rules lead to emergent segregation patterns akin to those observed in cities \cite{Schelling1971DynamicMO}.

Self-organizing systems are highly sensitive to initial conditions. This sensitivity results in stochastic emergent patterns. For example, the formation of fingerprints or animal skin patterns is driven by genetic signals, local cellular feedback, and mechanical constraints that result in unique patterns on each individual. Even slight variations in initial conditions can produce vastly different results. This property of complex systems poses a significant challenge to inverse problem-solving methods: it requires defining a loss metric that captures the equivalence of patterns with inherent variability.

\section{Inverse modelling in embedding space}
\label{sec:method}

\begin{figure}[!ht]
    \centering
    \begin{minipage}{0.85\textwidth}
    \sbox{\panelbox}{\includegraphics[width=0.98047\linewidth]{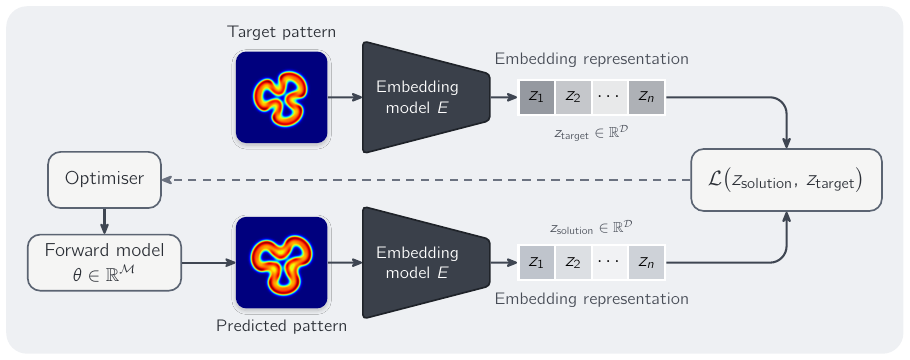}}%
    \leavevmode\rlap{\hspace*{0.00713\linewidth}\raisebox{\dimexpr\ht\panelbox-\height-0.0048\linewidth\relax}{\usefont{T1}{cmbr}{bx}{n}\fontsize{\the\dimexpr0.02572\linewidth\relax}{0pt}\selectfont a}}%
    \hspace*{0.01838\linewidth}\usebox{\panelbox}\hfill
    \end{minipage}

    \medskip
    \includegraphics[width=0.85\textwidth]{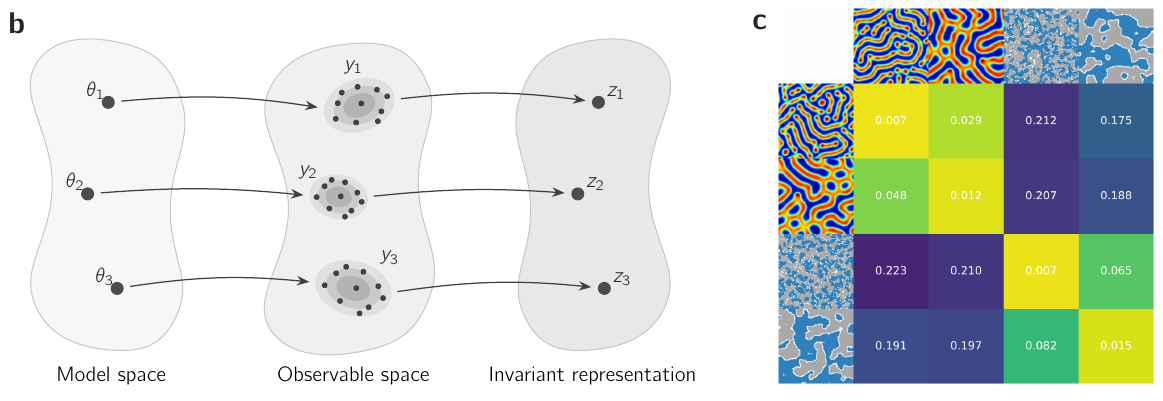}
    \caption{\textbf{Method overview and invariant representations.} (\textbf{a}): A target pattern is mapped onto an embedding space  \(z_{\text{target}} \in \mathbb{R}^\mathcal{D}\) using the embedding model $E$, where $\mathcal{D}$ is the embedding space dimension. A forward model, parametrised by \(\theta \in \mathbb{R}^\mathcal{M}\), where $\mathcal{M}$ is the degrees of freedom of the model, generates a pattern which is mapped to the same embedding space resulting in $z_{\text{solution}}$. The loss \(\mathcal{L}\) between the target and the candidate solution embedding guides the optimizer, iteratively updating \(\theta\) until convergence. By operating in embedding space, the method offers robustness against the inherent stochasticity of emergent patterns. (\textbf{b}): Diagram representing models parameters  $\theta_i$, generating observable patterns $y_i$, and the visual patterns being mapped to an embedding $z_i$. (\textbf{c}): Distance matrix between embedding representations $z$ of four patterns (two reaction-diffusion and two ABM segregation patterns). Notice that each of the patterns is a unique instantiation of the model. Similarity clusters appear, suggesting that the embedding model represents patterns with similar features as similar vectors. The values displayed are pairwise cosine distances $1 - \cos(z_i, z_j)$.}
    \label{fig:method}
\end{figure}

A common approach to solving inverse problems is to formulate them as optimization problems by defining a loss function that measures the discrepancy between target data and the optimizer predictions \cite{Vauhkonen2016Jul, Ye2019Mar}. However, the same parameters can generate patterns that share macroscopic features but differ pixel by pixel, making direct pattern comparison an unreliable measure of parameter recovery. To solve inverse problems with stochasticity in the observable space, a metric capable of capturing the feature-level similarities rather than exact pixel-level matches is needed. Embedding representations offer an effective solution to this challenge by encoding features and invariances of the patterns. The idea behind the proposed method is straightforward: map stochastic patterns onto an embedding space where perceptually similar patterns have similar vector representations. Then, use these invariant representations as the error signal to solve the inverse problem of recovering the original causal parameters that generate the stochastic patterns we seek to reconstruct. A visual summary is provided in Fig. \ref{fig:method}a.

The embedding model $E$ and loss function $\mathcal{L}$ are interchangeable. We evaluate four embeddings: Fréchet Wavelet, built from a multi-scale wavelet decomposition, and three learned embedding models, CLIP, Nomic and DINO; alongside domain-specific similarity metrics. Direct pattern MSE, which compares patterns pixel to pixel without an embedding model, serves as the baseline. Both the target and the predicted patterns are mapped onto the same embedding space, resulting in similar representations despite stochastic variations. An optimizer is then used to recover the original causal parameters, using the embedding distance between the target pattern and the predictions as the error signal.

The method is optimizer-agnostic: it can be used with gradient-based optimizers when the forward model and the embedding model are differentiable, or with black-box optimizers that do not require computing gradients. We use a black-box optimizer as the more general case, since it makes no assumptions about the differentiability of the forward model and permits application to physical systems where gradients are unavailable. We choose CMA-ES (Covariance Matrix Adaptation Evolution Strategy), a gradient-free population-based optimizer \cite{Hansen1996May}. Evolutionary strategies have been shown to find high-performing solutions in a variety of tasks \cite{Rios2013Jul, Belkhir2017Jul, Varelas2018Aug}--including in the context of inverse problems \cite{Bledsoe2011Oct, Grayver2016May}. The CMA-ES update is described in appendix Section \ref{sec:algorithm}.

The method operates as follows: \textbf{(1)} sample a set of parameters $\theta_i$ and random initial states $s_0 \sim \mathcal{S}$, \textbf{ (2)} generate the corresponding patterns $y_i$ through the forward model $F$, \textbf{(3)} use the embedding model to encode the resulting patterns into the embedding space $z$, \textbf{(4)} compute the loss between the target $z_{target}$ and the candidate solutions' embeddings $z_{solution}$, and \textbf{(5)} update the optimizer accordingly. Optimization maximizes the similarity between the two embeddings. By focusing on high-level visual similarities rather than pixel-level matches, the method is robust to stochastic variations, and is able to solve inverse problems across entire families of patterns with shared visual features--without requiring handcrafted loss functions. The method is detailed in appendix Algorithm \ref{alg:inverse_problem}.

\section{Experiments and Results}
In this section, we present the results demonstrating the method, solving inverse problems in embedding space, on three self-organizing systems: a reaction-diffusion system known as the Gray-Scott model \cite{Pearson1993Jul, Lee1993Jul}, a model of embryonic development \cite{cang2021multiscale}, and Schelling's agent-based model of segregation \cite{Schelling1971DynamicMO}. We chose these three distinct systems, both in terms of scientific domain (physics, biology and sociology) and the underlying mathematical model (PDE vs. ABM), to highlight the generality of the method. We first introduce the models, then formulate them as inverse problems, and subsequently present the results of applying the method with the wavelet embedding and the three learned embedding models, evaluating recovery against the ground-truth causal parameters. The results show that the method is encoder-agnostic. We compare it against the direct pattern MSE baseline, the same optimization without an embedding, and against handcrafted metrics that capture only some of the invariances of the patterns (Table \ref{table_results}). Finally, we evaluate the method on two non-simulated patterns: an ocellated lizard and a zebra skin pattern.

\begin{table}[H]
\centering
{\small
\setlength{\tabcolsep}{3.5pt}
\begin{tabular}{llcccc}
\hline
& Type & Reaction-Diffusion & Blastocyst & Schelling & z-score \\
\hline
\rowcolor{gray!20} Direct pattern MSE & Baseline & $0.158 \pm 0.082$ & $0.507 \pm 0.099$ & $0.200 \pm 0.129$ & $0.74$ \\
\hline
CLIP & Embedding & $0.038 \pm 0.021$ & $\mathbf{0.321} \pm 0.097$ & $0.044 \pm 0.033$ & $\mathbf{-0.99}$ \\
DINO & Embedding & $\mathbf{0.017} \pm 0.010$ & $0.423 \pm 0.044$ & $0.044 \pm 0.032$ & $-0.75$ \\
Nomic & Embedding & $0.042 \pm 0.022$ & $0.408 \pm 0.028$ & $0.054 \pm 0.041$ & $-0.64$ \\
Fréchet Wavelet & Wavelet embedding & $0.050 \pm 0.032$ & $0.333 \pm 0.087$ & $\mathbf{0.040} \pm 0.031$ & $-0.90$ \\
\hline
Dissimilarity index & Domain-specific & $0.222 \pm 0.060$ & $0.483 \pm 0.107$ & $\mathbf{0.040} \pm 0.028$ & $0.47$ \\
Edge density & Domain-specific & $0.111 \pm 0.045$ & $0.517 \pm 0.198$ & $0.265 \pm 0.226$ & $0.74$ \\
Orientation variance & Domain-specific & $0.126 \pm 0.072$ & $0.639 \pm 0.119$ & $0.299 \pm 0.239$ & $1.34$ \\
\hline
\end{tabular}}
\setlength{\abovecaptionskip}{7pt}
\caption{Mean absolute error (MAE) between the recovered parameters $\theta'$ and the ground-truth parameters $\theta$ (median ± MAD across ten runs) for ten runs per target pattern under three embedding metrics (CLIP, Nomic, DINO), Fréchet Wavelet, direct pattern (pixel-to-pixel) MSE and the domain-specific metrics. Direct pattern MSE is the baseline: the same optimization without an embedding model. Every row reports parameter-recovery error with respect to ground truth values, while the corresponding similarity metric guides optimization. Z-scores are computed by first standardizing each metric within each system $z_i = (x_i - \mu_i) / \sigma_i$ where $\mu_i$ and $\sigma_i$ are the mean and population standard deviation of the eight reported median MAEs within system $i$, then averaging these standardized values across the three systems. Baseline shown first; other rows grouped by type and ranked by unrounded z-score within each group. The lowest median MAE in each system and the lowest z-score are shown in bold.}
\label{table_results}
\end{table}

\subsection{Physical system: Reaction-Diffusion system}

Gray-Scott is a reaction-diffusion model used to study pattern formation in developmental systems \cite{Pearson1993Jul, Lee1993Jul}. It describes how spatially distributed concentrations of two interacting substances--or morphogens--evolve over time. Morphogens play a critical role in developmental biology by guiding cell differentiation and tissue patterning through their concentration gradients. $u$ represents a nutrient or precursor substance, while $v$ acts as an autocatalyst that promotes its own production while consuming $u$. The interaction between $u$ and $v$, combined with their ability to diffuse through space, leads to the emergence of complex patterns such as spots, stripes, or labyrinths. Through this interplay, the Gray-Scott model provides insights into how local reactions and diffusion contribute to the self-organization of biological structures. It is formalised as two partial differential equations:

\begin{minipage}{\textwidth}
\begin{tabular}{p{0.3\textwidth} p{0.65\textwidth}}
\[
\begin{aligned}
\frac{\partial u}{\partial t} &= D_u \nabla^2 u - uv^2 + f(1 - u) \\
\frac{\partial v}{\partial t} &= D_v \nabla^2 v + uv^2 - (f + k)v
\end{aligned}
\]
&
\begin{itemize}
    \item[-] $u$ and $v$ are the chemical concentrations.
    \item[-] $D_u$ and $D_v$ are the diffusion rates for $u$ and $v$.
    \item[-] $f$ is the feed rate of $u$.
    \item[-] $k$ is the kill rate of $v$.
    \item[-] $\nabla^2$ is the Laplacian representing diffusion.
\end{itemize}
\end{tabular}
\label{RD_equations}
\end{minipage}

By convention, the $u$ field is treated as a latent variable, while the $v$ field represents the morphogens interpreted as visible patterns. 

\subsubsection{Results Reaction-Diffusion}

To find the unknown parameters for the reaction-diffusion system, we follow the algorithm described in appendix Algorithm \ref{alg:inverse_problem}. At each optimization run, we generate a target pattern for a set of parameters $\{D_u, D_v, k, f\}_{target} \in \mathbb{R}^4$ by simulating Gray-Scott partial differential equations \ref{RD_equations} from a random initial state $s_0 \sim \mathcal{S}$ for 1000 steps. The resulting pattern is embedded onto $z_{target}$ with the selected embedding model. $z_{target}$ serves as the target embedding for the loss throughout the optimization process. Then, the optimizer samples a population of candidate solutions, $\{D_u, D_v, k, f\}_i \in \mathbb{R}^4$, used to simulate the dynamics of reaction-diffusion. The generated patterns $y_i$ are then passed through the embedding model, resulting in $z_i$. We compare each  $z_i$ against the target embedding  $z_{target}$ using cosine similarity $cos(z_i, z_{target})$. The optimization is iteratively run until it converges to a solution that maximizes the cosine similarity with respect to the target embedding  $z_{target}$. 

\begin{figure}[!ht]
    \centering
    \newcommand*{\gstile}[1]{\includegraphics[width=0.0785\textwidth, trim=#1, clip]{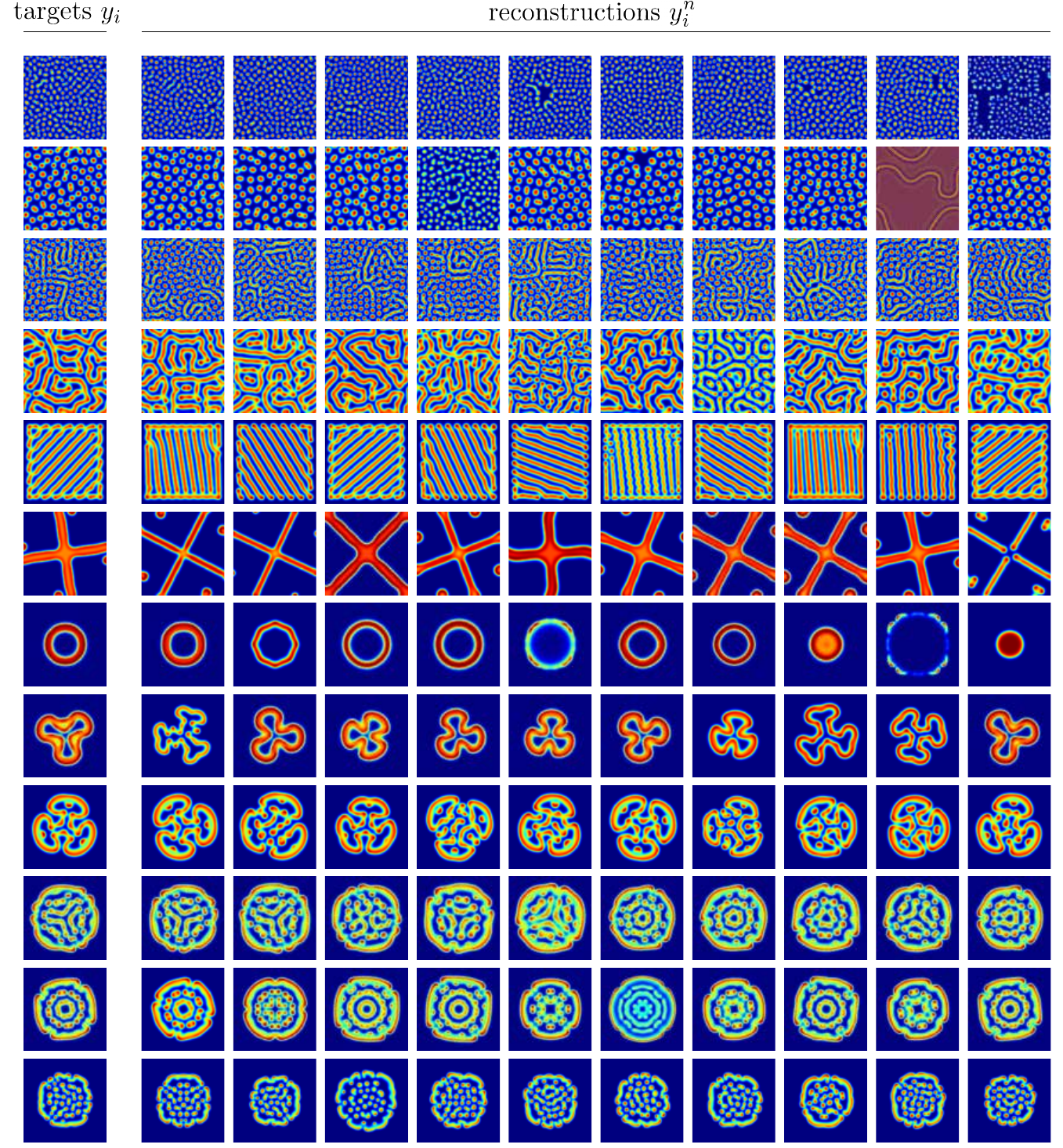}}%
    \newcommand*{\gsheader}[1]{{\small #1}\par\smallskip\hrule\smallskip}%
    \gsheader{targets $y_i$}
    \gstile{14.40 617.69 585.51 33.92}\hspace{0.005\textwidth}%
    \gstile{14.40 561.83 585.51 89.78}\hspace{0.005\textwidth}%
    \gstile{14.40 505.97 585.51 145.64}\hspace{0.005\textwidth}%
    \gstile{14.40 450.10 585.51 201.51}\hspace{0.005\textwidth}%
    \gstile{14.40 394.24 585.51 257.37}\hspace{0.005\textwidth}%
    \gstile{14.40 338.38 585.51 313.23}\hspace{0.005\textwidth}%
    \gstile{14.40 282.51 585.51 369.10}\hspace{0.005\textwidth}%
    \gstile{14.40 226.65 585.51 424.96}\hspace{0.005\textwidth}%
    \gstile{14.40 170.79 585.51 480.82}\hspace{0.005\textwidth}%
    \gstile{14.40 114.92 585.51 536.69}\hspace{0.005\textwidth}%
    \gstile{14.40 59.06 585.51 592.55}\hspace{0.005\textwidth}%
    \gstile{14.40 3.20 585.51 648.41}%

    \medskip
    \gsheader{reconstructions $y_i'$}
    \gstile{86.67 617.69 513.24 33.92}\hspace{0.005\textwidth}%
    \gstile{86.67 561.83 513.24 89.78}\hspace{0.005\textwidth}%
    \gstile{86.67 505.97 513.24 145.64}\hspace{0.005\textwidth}%
    \gstile{86.67 450.10 513.24 201.51}\hspace{0.005\textwidth}%
    \gstile{86.67 394.24 513.24 257.37}\hspace{0.005\textwidth}%
    \gstile{86.67 338.38 513.24 313.23}\hspace{0.005\textwidth}%
    \gstile{86.67 282.51 513.24 369.10}\hspace{0.005\textwidth}%
    \gstile{86.67 226.65 513.24 424.96}\hspace{0.005\textwidth}%
    \gstile{86.67 170.79 513.24 480.82}\hspace{0.005\textwidth}%
    \gstile{86.67 114.92 513.24 536.69}\hspace{0.005\textwidth}%
    \gstile{86.67 59.06 513.24 592.55}\hspace{0.005\textwidth}%
    \gstile{86.67 3.20 513.24 648.41}%
    \caption{\textbf{Results for Gray-Scott reaction-diffusion system.} Twelve target patterns $y_i$ (\textit{top row}) and one reconstruction $y_i'$ per target (\textit{bottom row}), generated from the parameters recovered by the method. Reconstructions for all 10 independent training runs, each with a different random seed, are shown in appendix Fig.~\ref{fig:RD_grid}.}
    \label{fig:RD_examples}
\end{figure}

Results for the twelve target patterns are shown in Fig. \ref{fig:RD_examples}. 
Results for 10 independent training runs with different random seeds on the 12 target patterns are shown in appendix Fig. \ref{fig:RD_grid}. Across the twelve target patterns, the embedding models reach a parameter-recovery MAE of 0.017--0.042, compared with 0.158 for the direct pattern MSE baseline (Table \ref{table_results}).

\subsection{Biological system: Embryonic development}

A blastocyst is the early-stage of the mammalian embryo. It is composed of three main components, an outer cellular layer, the trophoblast, which later in development becomes the placenta, and an inner cellular cluster, the embryoblast that develops into the organism itself, and a layer, the hypoblast, separating the embryoblast from the internal cavity; the cellular arrangement of the embryo at the blastocyst stage is shown in Fig. \ref{fig:blastocyst_allpatterns} (\textit{left column}). Due to the stochastic nature of morphogenesis, the exact position where the stem cells differentiate the embryoblast varies with each developmental instance.

The blastocyst simulation is based on \citet{cang2021multiscale} Cellular Potts Model (CPM), where the gene regulatory network model defined as a set of stochastic differential equations guide the cells' differentiation. Unlike in the reaction-diffusion experiments, the blastocyst has a single stochastic target pattern, representing the observed embryogenic developmental process in mammalian organisms \cite{biggers1988mammalian, cang2021multiscale}. The simulation is performed using Morpheus library and based on  \citet{Morpheus2025Apr}'s implementation.

\vspace{-1mm}
\subsubsection{Results - Embryonic development}
\begin{figure}[htbp]
   \centering
    \newcommand*{\bcheader}[1]{{\small #1}\par\smallskip\hrule\smallskip}%
    \begin{minipage}[b]{0.10373\textwidth}
    \centering
    \bcheader{target $y$}
    \includegraphics[width=\linewidth, trim=18.3 0.9 571.03 24, clip]{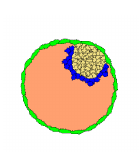}
    \end{minipage}\hspace{0.00937\textwidth}%
    \begin{minipage}[b]{0.88690\textwidth}
    \centering
    \bcheader{reconstructions $y^n$}
    \includegraphics[width=\linewidth, trim=88 0.9 0 24, clip]{figures/blastocyst_result_grid.pdf}
    \end{minipage}
    \vspace{-4mm}
\caption{\textbf{Results for blastocyst embryonic developmental model.} A single target pattern $y$ \textit{(left column)}, simulated based on a Cellular Potts model \cite{Morpheus2025Apr}. The developmental process is implemented as a set of stochastic differential equations  \cite{cang2021multiscale, Morpheus2025Apr}. The reconstructions columns show the patterns $y^n$ recovered by our method using CLIP embeddings for 10 independent training runs. Non-uniqueness in the forward system allows distinct parameter sets to produce successful developmental patterns, so large parameter-recovery MAE can coexist with successful pattern reconstruction.}
    \label{fig:blastocyst_allpatterns}
\end{figure}

In accordance with the embryonic development literature \cite{cang2021multiscale}, a successful developmental process is one that leads to a cellular arrangement where the inner cell mass, or embryoblast, adheres to the outer layer, or trophoblast, while having a thin endodermic layer, or hypoblast, facing towards the inner cavity. An instance of this developmental pattern is shown in Fig. \ref{fig:blastocyst_allpatterns} (\textit{target $y$}). The inverse problem consists in finding a set of four Hill coefficients that modulate the expression of several transcription factors and proteins, acting as a complex genetic switch that guides development. Next to the target pattern, Fig. \ref{fig:blastocyst_allpatterns} shows the outcome of ten independent optimization runs.  
The method recovers successful developmental patterns in seven out of ten runs, compared with one out of ten using the direct pattern MSE baseline for the same optimization budget.
Results for all embedding models and metrics are shown in Table \ref{table_results}.

\vspace{-1mm}

\vspace{-3mm}
\subsection{Social system: Schelling's Model of Segregation}
\vspace{-1mm}
The third system we use to demonstrate the method is Schelling's segregation model \cite{Schelling1971DynamicMO}, a classic computational model used to study spatial pattern emergence in social systems. It illustrates how individual preferences for neighbourhood composition can lead to large-scale segregation. Schelling's model has been highly influential in computational social science, contributing to the understanding of segregation and its implications for urban planning and social dynamics. Schelling's model has historical significance in computational social science, as it was one of the first to show how simple local rules can produce complex emergent behaviours, highlighting the potential of agent-based models for social systems \cite{Paczuski1999Jun,clark2008understanding}. For this latter reason, we chose it to demonstrate our method.

Our implementation consists of two classes of agents, randomly distributed on a spatial grid. Each agent evaluates its satisfaction based on the proportion of similar neighbours in its local neighbourhood. Unsatisfied agents with their current location will asynchronously relocate to a random location to improve their satisfaction. Schelling's model leads to the emergence of complex segregation patterns that match those observed in urban environments.

\vspace{-2mm}
\begin{minipage}{\textwidth}
\begin{tabular}{p{0.3\textwidth} p{0.65\textwidth}}
\[
S_i = 
\begin{cases} 
1 & \text{if } \frac{N_{\text{similar}}}{N_{\text{total}}} \geq T \\
0 & \text{otherwise}
\end{cases}
\]
&
\begin{itemize}
    \item[-] $S_i$ is the satisfaction of agent $i$.
    \item[-] $N_{\text{similar}}$ is the number of similar neighbours.
    \item[-] $N_{\text{total}}$ is the total number of neighbours.
    \item[-] $T$ is the agent's tolerance threshold, $T \in [0,1]$.
\end{itemize}
\end{tabular}
\label{SH_equations}
\end{minipage}

The emergent patterns, while differing at the microscopic level, exhibit consistent macroscopic structures and statistical properties. It is these invariances of the macroscopic patterns that our method exploits to find the unknown threshold.

\subsubsection{Results - Schelling Segregation Model}
\begin{figure}[!ht]
    \centering
    \newcommand*{\shtile}[1]{\includegraphics[width=0.1206\linewidth, trim=#1, clip]{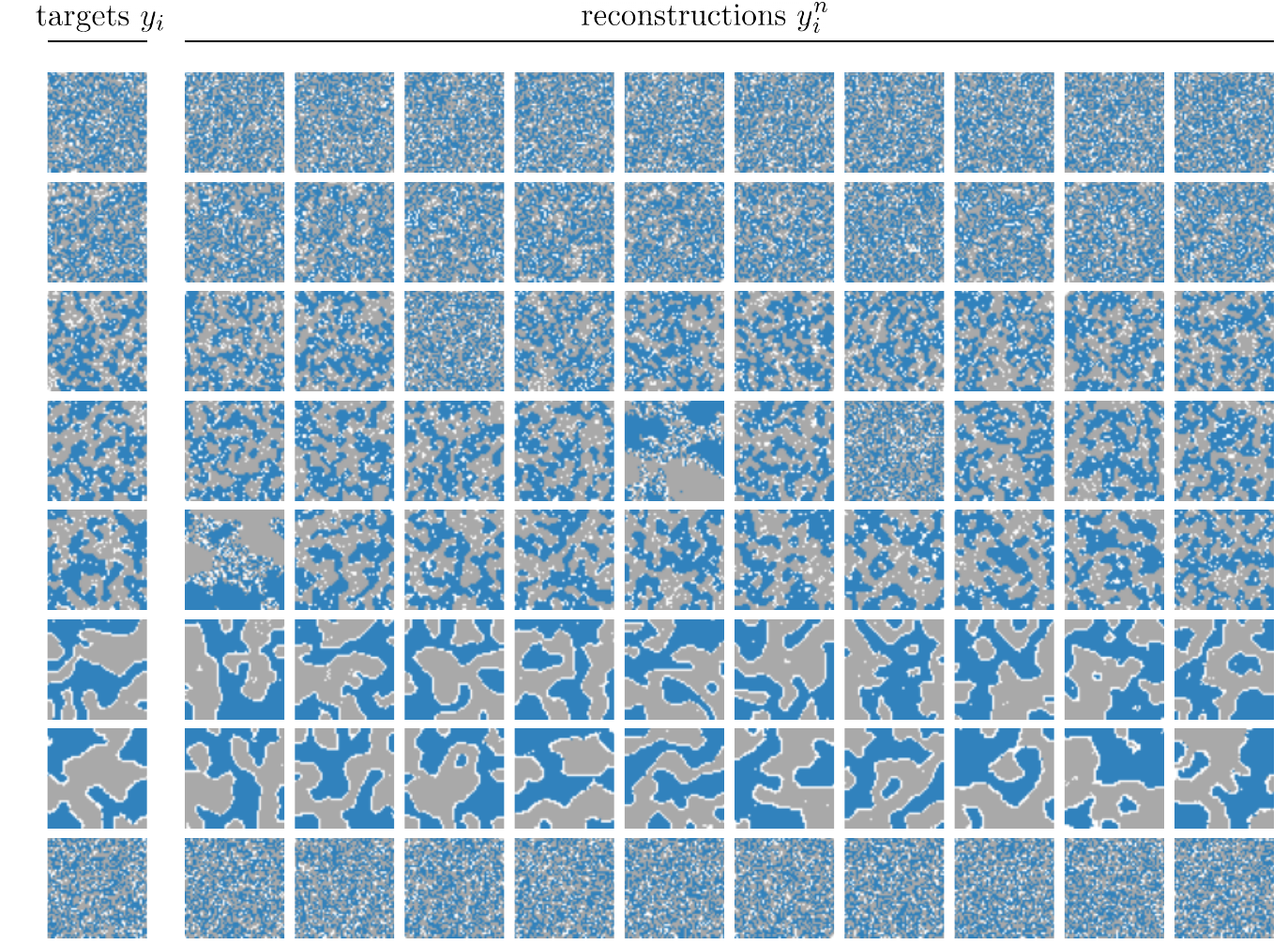}}%
    \newcommand*{\shheader}[1]{{\small #1}\par\smallskip\hrule\smallskip}%
    \begin{minipage}{\textwidth}
    \centering
    \shheader{targets $y_i$}
    \shtile{24.32 395.14 583.59 36.94}\hspace{0.005\linewidth}%
    \shtile{24.32 339.19 583.59 92.89}\hspace{0.005\linewidth}%
    \shtile{24.32 283.24 583.59 148.84}\hspace{0.005\linewidth}%
    \shtile{24.32 227.29 583.59 204.79}\hspace{0.005\linewidth}%
    \shtile{24.32 171.34 583.59 260.74}\hspace{0.005\linewidth}%
    \shtile{24.32 115.39 583.59 316.69}\hspace{0.005\linewidth}%
    \shtile{24.32 59.45 583.59 372.63}\hspace{0.005\linewidth}%
    \shtile{24.32 3.50 583.59 428.58}%

    \medskip
    \shheader{reconstructions $y_i'$}
    \shtile{94.58 395.14 513.33 36.94}\hspace{0.005\linewidth}%
    \shtile{94.58 339.19 513.33 92.89}\hspace{0.005\linewidth}%
    \shtile{94.58 283.24 513.33 148.84}\hspace{0.005\linewidth}%
    \shtile{94.58 227.29 513.33 204.79}\hspace{0.005\linewidth}%
    \shtile{94.58 171.34 513.33 260.74}\hspace{0.005\linewidth}%
    \shtile{94.58 115.39 513.33 316.69}\hspace{0.005\linewidth}%
    \shtile{94.58 59.45 513.33 372.63}\hspace{0.005\linewidth}%
    \shtile{94.58 3.50 513.33 428.58}%
    \end{minipage}
    \caption{\textbf{Results for Schelling segregation model.} Eight target patterns $y_i$ (\textit{top row}) for \emph{tolerance} parameter $\mathbf{T} \in \{ 0.1\mathbf{i} \mid \mathbf{i} \in \{1,2,\dots,8\} \}$ (\textit{left to right}), and one reconstruction $y_i'$ per target (\textit{bottom row}), generated from the parameters recovered by the method. \normalfont For small values of $T$, agents converge with minimal migration as tolerance threshold is quickly satisfied, in middle range of $T$ (3-6) segregation patterns become more pronounced, while for large values of $T$ the system does not converge as threshold is unattainable. Reconstructions for all 10 independent training runs, each with a different random seed, are shown in appendix Fig.~\ref{fig:SCHELLING_grid}.}
    \label{fig:SH_results}
\end{figure}

In this experiment, the same optimization methodology is applied to find the tolerance $T$ from different Schelling segregation patterns. We use a grid size of 100 by 100, and an occupation density of 90\%. The results of the experiment are presented in Figure  \ref{fig:SH_results}. The method successfully finds the inverse parameters. Over 10 independent training runs, the embedding models reach a parameter-recovery MAE of 0.044--0.054, compared with 0.200 for the direct pattern MSE baseline (Table \ref{table_results}). The patterns resulting from each training run are shown in the appendix Fig. \ref{fig:SCHELLING_grid}.

\vspace{-2mm}
\subsection{Comparison of representations and similarity metrics}
\vspace{-1mm}
Our evaluation demonstrates that the method works across different systems without having to design or learn system-specific similarity metrics. We demonstrate its encoder-agnostic design using a wavelet embedding and multiple learned embedding models, and compare against the direct pattern MSE baseline and handcrafted metrics to illustrate the benefit of representations that can be reused across systems.

The experiment results are shown in Table \ref{table_results} as mean absolute errors (MAE) between the recovered parameters and the ground-truth parameters, using each metric as the optimization error signal; variability is reported using median absolute deviation (MAD) given non-normal error distributions. We compare eight representations and similarity metrics: Fréchet Wavelet, an embedding built from a wavelet packet decomposition \cite{Veeramacheneni2023Dec}; three learned embedding models, CLIP, Nomic \cite{Nussbaum2024Jun} and DINO \cite{caron2021emerging}; the direct pattern MSE baseline; and three domain-specific metrics: dissimilarity index, which measures spatial clustering patterns in agent-based models; edge density, quantifying boundary structures; and orientation variance, capturing directional pattern features. The remaining optimization settings are fixed within each system, with a population size and number of generations of 128 and 32 for reaction-diffusion, 16 and 20 for blastocyst, and 16 and 25 for Schelling.

All three embedding models achieve lower parameter-recovery MAE than the direct pattern MSE baseline in every system, the direct comparison commonly used in optimization-based inverse problems \cite{Vauhkonen2016Jul, Ye2019Mar}. CLIP and Fréchet Wavelet achieve the lowest aggregate z-scores (-0.99 and -0.90), showing that both represent the diverse patterns of the three systems in a way that is informative for parameter recovery. DINO (-0.75) and Nomic (-0.64) also achieve negative aggregate z-scores, indicating lower-than-average parameter-recovery error across systems. No single embedding model performs best in every system: DINO achieves the best MAE on reaction-diffusion and CLIP on blastocyst. Unsurprisingly, dissimilarity index, a domain-specific metric designed to measure spatial clustering, achieves the lowest error on Schelling (0.040), and Fréchet Wavelet matches its performance despite being a general metric. The embedding models perform similarly well on Schelling (0.044--0.054) and generalize across all three systems, whereas dissimilarity index performs poorly on the other patterns (z-score: 0.47). The other domain-specific metrics, edge density and orientation variance, perform substantially worse overall (z-scores: 0.74 and 1.34).

\subsection{Natural patterns}

In this section, we evaluate the method on natural patterns, rather than simulated ones as in the previous experiments. First, we use the skin pattern of an ocellated lizard from \citet{Manukyan2017Apr}, who attributed it to a cellular automaton driven by an underlying reaction-diffusion process. We apply our method to find a set of Gray-Scott reaction-diffusion model parameters capable of reconstructing the lizard patterns. Here, Gray-Scott serves as a conjectured model of pattern formation, without assuming knowledge of the actual dynamics generating the lizard's skin pattern.  The target and reconstructed patterns are shown in Fig. \ref{fig:natural}a.

\paragraph{Guiding Model Testing}
Identifying unknown parameters underlying natural patterns is more challenging, as there is no guarantee that the chosen model can generate the target patterns. When the underlying dynamics are unknown, the search can extend from the parameters of a conjectured model to the model's functional form \cite{brunton2016discovering}, using the same embedding loss to compare generated and observed patterns. Symbolic regression provides one approach to searching over functional forms \cite{cranmer2023interpretable}. However, the space of possible models is extremely large, and we often have an intuition about the model. In such cases, we can conjecture a minimal model form that might be driving the process that generates the target pattern. We can then apply the inverse method to tune its free parameters to reconstruct the natural target. If the model's expressivity is insufficient to capture the pattern, we can iteratively complexify the model. This approach allows us to preserve a mechanistic understanding of the underlying process until the model captures the target pattern.

In this second experiment with natural patterns, we aim at finding the reaction-diffusion parameters that lead to a zebra skin pattern \cite{Lindner2024May}. In our initial attempt at generating a zebra-like pattern, we found that the standard Gray-Scott reaction-diffusion model failed to produce matching results. This suggests that the underlying mechanism responsible for the zebra pattern differs from the assumed model. The Gray-Scott model is isotropic, meaning it treats diffusion equally in all directions, which makes it challenging to generate stripe patterns. To overcome this limitation, we introduced anisotropy into the partial differential equations by allowing the diffusion coefficients to vary directionally:

\begin{figure}[!ht]
    \centering
    \newcommand*{\natcol}[2]{\begin{minipage}[t]{0.16\textwidth}\centering{\small #1}\par\smallskip\hrule\medskip#2\end{minipage}}%
    \panel{a}{\natcol{target $y$}{\includegraphics[width=\linewidth, trim=31 141 770 11, clip]{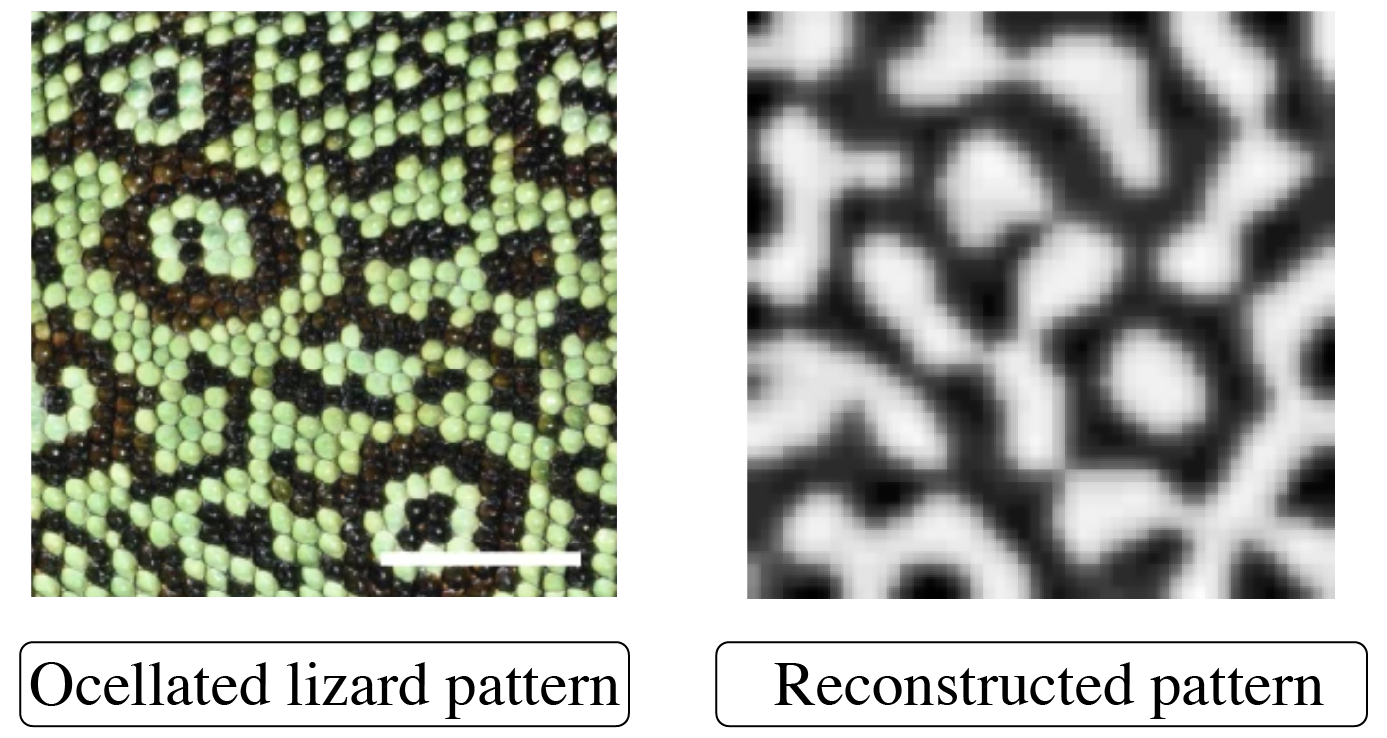}}}\hspace{0.03\textwidth}%
    \natcol{reconstruction $y'$}{\includegraphics[width=\linewidth, trim=749 141 52 11, clip]{figures/gecco.png}}\hspace{0.08\textwidth}%
    \panel{b}{\natcol{target $y$}{\includegraphics[width=\linewidth, trim=20.64 125.54 746.49 9.84, clip]{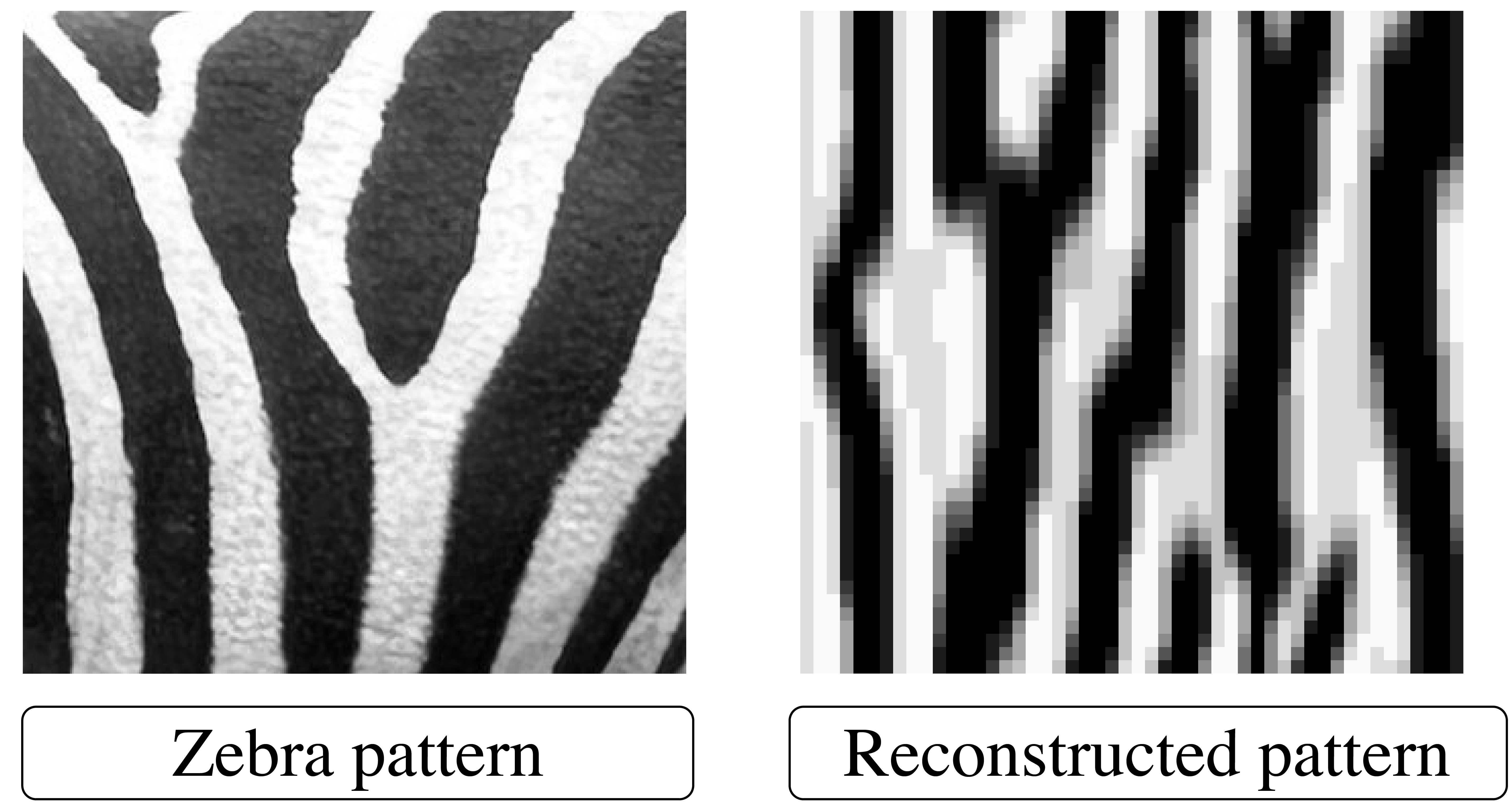}}}\hspace{0.03\textwidth}%
    \natcol{reconstruction $y'$}{\includegraphics[width=\linewidth, trim=726.09 125.54 41.29 9.84, clip]{figures/zebra.png}}
    \caption{\textbf{Reconstruction of natural patterns.} (\textbf{a}): Ocellated lizard pattern. \textit{Left:} Photograph of an ocellated lizard from \citet{Manukyan2017Apr}, white bar indicates 5mm scale. \textit{Right:} Reconstructed pattern by the set of reaction-diffusion parameters found by the method. (\textbf{b}): Zebra pattern. \textit{Left:} Photograph of a zebra skin pattern \cite{Lindner2024May}. \textit{Right:} Pattern generated by the anisotropic reaction-diffusion model for the parameters found by the method.}
    \label{fig:natural}
\end{figure}

\begin{minipage}{\textwidth}
\begin{minipage}[c]{0.45\textwidth}
\centering
$\displaystyle
\begin{aligned}
\frac{\partial u}{\partial t} &= D_{u_x} \frac{\partial^2 u}{\partial x^2} + D_{u_y} \frac{\partial^2 u}{\partial y^2} - u v^2 + f (1 - u) \\
\frac{\partial v}{\partial t} &= D_{v_x} \frac{\partial^2 v}{\partial x^2} + D_{v_y} \frac{\partial^2 v}{\partial y^2} + u v^2 - (f + k) v
\end{aligned}$
\end{minipage}\hspace{0.05\textwidth}%
\begin{minipage}[c]{0.45\textwidth}
\begin{itemize}
\item[-] $D_{u_x}$ and $D_{u_y}$ are the diffusion rates for $u$ in the $x$- and $y$-directions.
    \item[-] $D_{v_x}$ and $D_{v_y}$ are the diffusion rates for $v$ in the $x$- and $y$-directions.
\end{itemize}
\end{minipage}
\end{minipage}

All other terms remain the same as in the standard Gray-Scott model. This anisotropic modification allows for spatial variation in the diffusion dynamics, introducing a directional bias that can generate the stripe patterns required for the zebra pattern. As a result, the anisotropic Gray-Scott model successfully produces the desired pattern, as shown in Fig. \ref{fig:natural}b.

Along with the other approaches discussed in Section \ref{inverse problems}, the expressivity bottleneck of the model exemplifies how our method can guide the search for mechanistic models. While the method can simply fail due to the optimization being challenging, by iteratively expanding a conjectured model until it achieves an expressivity level compatible with observed data, we can use the method to potentially identify missing mechanisms and refine our understanding of pattern formation in nature.

\section{Conclusion} 

In this work, we have presented a method for solving stochastic inverse problems in embedding space, demonstrated on self-organizing systems--without the need to handcraft metric functions to capture the visual similarities between patterns. The key idea is to shift the optimization from direct pattern space to an embedding space where perceptually similar patterns are mapped to nearby points, allowing the recovery of the original causal parameters, even those resulting in stochastic patterns. We have shown that such an embedding can be built from a multi-scale wavelet decomposition or taken from visual embedding models trained on large image datasets. The method addresses stochasticity in the observable space without having to design or learn a system-specific similarity metric or summary statistic. This avoids the system-specific summary construction or learning used in ABC approaches \cite{beaumont2002approximate, jiang2017learning}, while enabling parameter recovery across three distinct self-organizing systems.

We have demonstrated the method on three distinct self-organizing systems spanning physics, biology and computational sociology: a reaction-diffusion system, a developmental model of embryogenesis, and an agent-based segregation model, all exhibiting stochasticity in the observation space. For the reaction-diffusion model, we have shown that the approach can be applied to real biological patterns: an ocellated lizard and a zebra skin pattern. Finally, we have illustrated how the framework can guide model testing: when a conjectured model fails to generate the target, it suggests that the model's expressivity is insufficient and extra terms are needed, a practical path for iterative refinement in mechanistic modeling. In summary, the method provides a simple yet effective technique for theorists and experimentalists to investigate and control self-organizing stochastic systems.

\newpage

\bibliographystyle{unsrtnat}
\bibliography{refs}

\newpage

\section{Appendix}

\subsection{Inverse modeling algorithm}
\label{sec:algorithm}
Algorithm \ref{alg:inverse_problem} details the method described in Section \ref{sec:method}. At each iteration, the optimizer samples candidate parameters $\theta_i$, the forward model $F$ generates the corresponding patterns from a random initial state, the embedding model $E$ maps the target and candidate patterns onto the embedding space, and the loss $\mathcal{L}$ between their embeddings is used to update the optimizer.

\begin{algorithm}[H]
\caption{Inverse modeling through embedding representation}
\label{alg:inverse_problem}
\begin{algorithmic}[1]
\Require Forward model $F$, State space $S$, Loss function $\mathcal{L}$,  Embedding model $E$, optimizer
\State Choose target pattern $y_{target}$
\While{not converged}
    \State Sample candidate solutions $\theta_{i}$ from optimizer 
    \State Sample an initial configuration $s_0 \in S$
    \State Run simulations: $y_{\text{solution}} \gets F(\theta_i, s_0 )$
    \State Map patterns to embedding space:
    \State \hspace{1em} $z_{\text{target}} \gets E(y_{\text{target}})$
    \State \hspace{1em} $z_{\text{solution}} \gets E(y_{\text{solution}})$
    \State Compute loss: $loss_i \gets \mathcal{L}(z_{\text{solution}}, z_{\text{target}})$
    \State Update optimizer state with $loss_i$
\EndWhile
\State \Return $\theta = \arg\min_{\theta} \text{loss}_i$
\end{algorithmic}
\end{algorithm}

\paragraph{Optimizer.} CMA-ES operates by iteratively updating the covariance matrix of a multivariate normal distribution at each generation, so that candidate solutions $\theta_i$ are increasingly guided towards promising regions of the search space. Solutions are iteratively sampled from a Gaussian distribution centred on the current mean with the covariance defining the search distribution:
$ \boldsymbol{\theta}_i = \mathbf{m} + \sigma \mathbf{C}^{1/2} \boldsymbol{\epsilon}$ with $\boldsymbol{\epsilon} \sim \mathcal{N}(\mathbf{0}, \mathbf{I})$ where $\mathbf{m}$ is the mean, $\sigma$ is the step-size, $\mathbf{C}$ is the covariance matrix and $\mathbf{I}$ is the identity matrix. The mean $\mathbf{m}$ is updated as a weighted sum of selected solutions $
\mathbf{m}_{\text{new}} = \sum_{k=1}^\mu w_k \boldsymbol{\theta}_k$ where $\mu$ is the number of selected solutions, and $w_k$ are the weights based on solution rank.

\subsection{Implementation details}
\paragraph{Hardware details.} Experiments have been run on a single machine with a \textit{AMD Ryzen Threadripper 3990X} CPU and a \textit{NVIDIA GeForce RTX 4090} GPU.

\paragraph{Reproducibility.} The anonymised codebase including the algorithm implementation and scripts used to run the reported experiments can be found in
\url{https://github.com/enajx/InverseStochastic/}.

\subsection{Additional results for each system}
\begin{figure}[htbp]
   \centering
    \includegraphics[width=\textwidth]{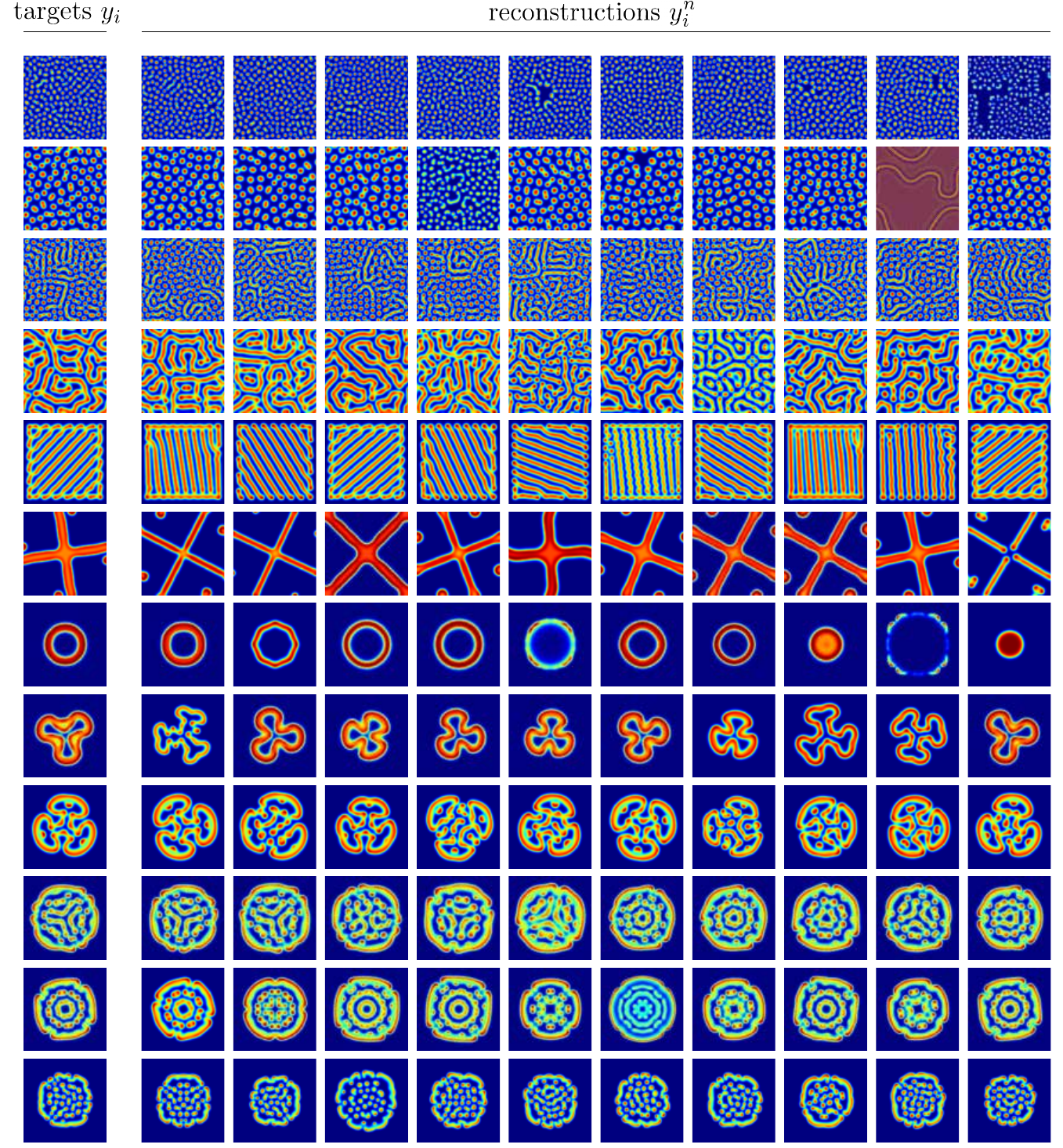}
\caption{\textbf{Results for Gray-Scott model.} Twelve target patterns $y_i$ \textit{(left column)}, each produced by simulating Gray-Scott partial differential equations \ref{RD_equations} for different parameters $\theta_i$ and random initial states $s_{0i}$. The reconstructions columns show the patterns $y_i^n$ recovered by our method for 10 independent training runs.}
\label{fig:RD_grid}
\end{figure}

\begin{figure}[htbp]
   \centering
    \includegraphics[width=\textwidth]{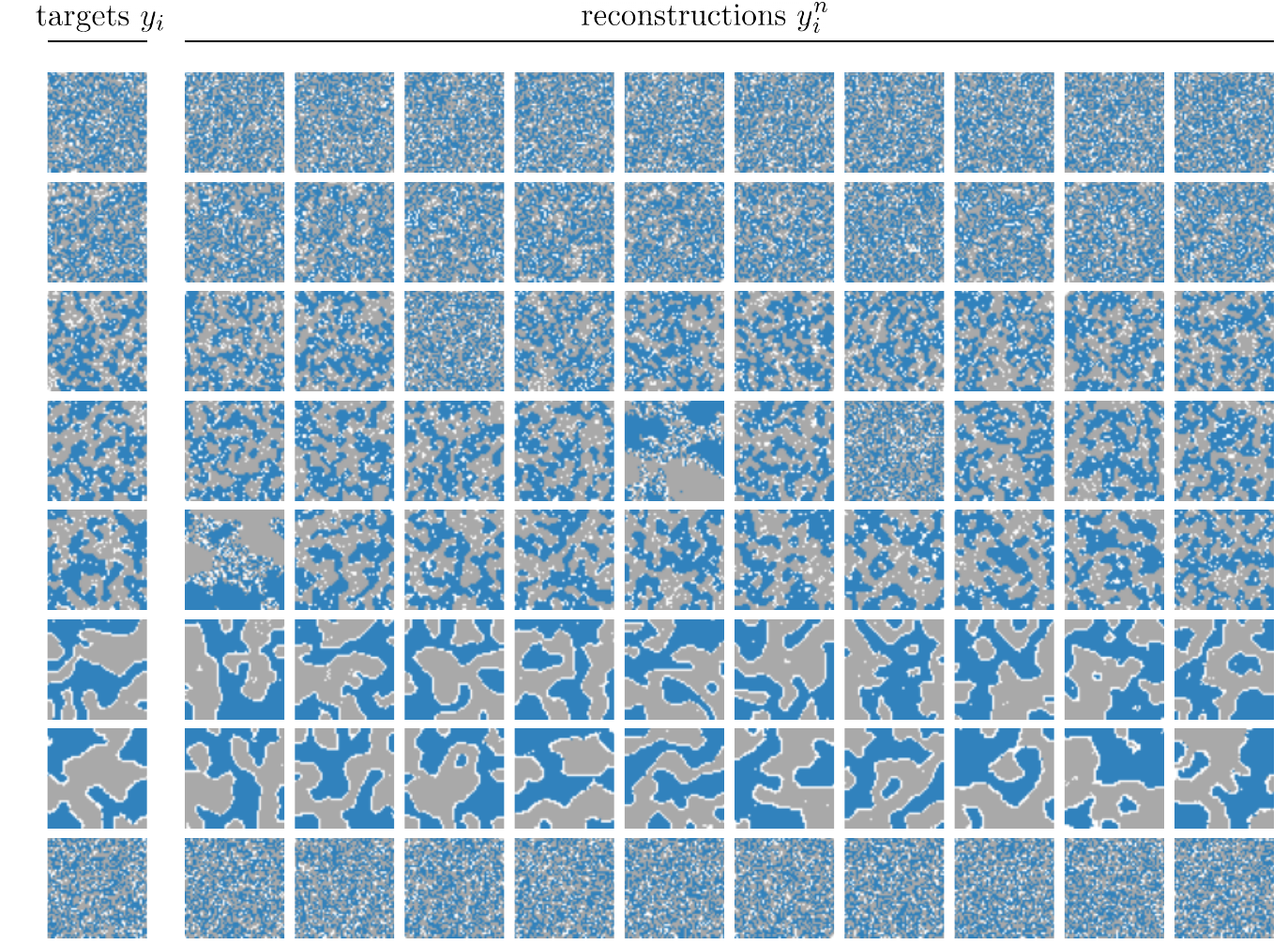}
\caption{\textbf{Results for Schelling's segregation model.} Eight target patterns $y_i$ \textit{(left column)}, each produced by simulating Schelling ABM model of segregation \ref{SH_equations} for different parameters $\theta_i$ and random initial states $s_{0i}$. The reconstructions columns show the patterns $y_i^n$ recovered by our method for 10 independent training runs.}
    \label{fig:SCHELLING_grid}
\end{figure}

Figs. \ref{fig:RD_grid} and \ref{fig:SCHELLING_grid} show the patterns recovered in 10 independent training runs for the Gray-Scott and Schelling models, respectively.

\newpage

\newpage

\clearpage

\end{document}